\documentstyle[12pt,epsf]{article}

\def\fo{\hbox{{1}\kern-.25em\hbox{l}}}
\def\fnote#1#2{\begingroup\def\thefootnote{#1}\footnote{#2}\addtocounter
{footnote}{-1}\endgroup}

\renewcommand{\thefootnote}{\fnsymbol{footnote}}

\def\beq{\begin{equation}}
\def\eeq{\end{equation}}
\def\eq{\end{equation}}
\def\to{\rightarrow}

\def\EmissT{\not \! \!  E_{T}}

\newcommand{\newc}{\newcommand}
 
\newc{\eegg}{e^+e^-\gamma\gamma}
\newc{\mmgg}{\mu \mu \gamma\gamma}
\newc{\ttgg}{\tau \tau \gamma\gamma}

\newc{\leplep}{l^+l^-}
\newc{\llgg}{l^+l^- \gamma \gamma}
\newc{\lllgg}{l^+l^-l^{\prime \pm} \gamma \gamma}
\newc{\ljjgg}{l^{\pm}jj \gamma \gamma}
\newc{\lljjgg}{l^+l^-jj \gamma \gamma}

\newc{\eeggE}{ee \gamma \gamma + \EmissT}
\newc{\llggE}{l^+l^- \gamma \gamma + \EmissT}

\newc{\gag}{\gamma\gamma}
\newc{\lgg}{l^{\pm} \gamma \gamma}

\newc{\jjgg}{jj \gamma \gamma}
\newc{\jjjjgg}{4j \gamma \gamma}

\newc{\gsim}{\lower.7ex\hbox{$\;\stackrel{\textstyle>}{\sim}\;$}}
\newc{\lsim}{\lower.7ex\hbox{$\;\stackrel{\textstyle<}{\sim}\;$}}
\newc{\ie}{{\it i.e.}}		
\newc{\etal}{{\it et al.}}
\newc{\eg}{{\it e.g.}}		
\newc{\kev}{\hbox{\rm\,keV}}		
\newc{\mev}{\hbox{\rm\,MeV}}		
\newc{\gev}{\hbox{\rm\,GeV}}		
\newc{\tev}{\hbox{\rm\,TeV}}
\newc{\xpb}{\hbox{\rm\, pb}}
\newc{\xfb}{\hbox{\rm\, fb}}

\newc{\mtop}{m_t}
\newc{\mbot}{m_b}
\newc{\mz}{m_Z}
\newc{\mw}{m_W}
\newc{\alphasmz}{\alpha_s(m_Z^2)}
\newc{\swsq}{\sin^2\theta_W}
\newc{\tw}{\tan\theta_W}
\newc{\cw}{\cos\theta_W}
\newc{\sw}{\sin\theta_W}
\newc{\BR}{\hbox{\rm BR}}
\newc{\zbb}{Z\to b\bar}
\newc{\Gb}{\Gamma (Z\to b\bar b)}
\newc{\Gh}{\Gamma (Z\to \hbox{\rm hadrons})}
\newc{\rbsm}{R_b^\hbox{\rm sm}}
\newc{\rbsusy}{R_b^\hbox{\rm susy}}
\newc{\drb}{\delta R_b}

\newc{\tbeta}{\tan\beta}
\newc{\uL}{{\tilde u_L}}
\newc{\uR}{{\tilde u_R}}
\newc{\cL}{{\tilde c_L}}
\newc{\cR}{{\tilde c_R}}
\newc{\tL}{{\tilde t_L}}
\newc{\tR}{{\tilde t_R}}
\newc{\dL}{{\tilde d_L}}
\newc{\dR}{{\tilde d_R}}
\newc{\sL}{{\tilde s_L}}
\newc{\sR}{{\tilde s_R}}
\newc{\bL}{{\tilde b_L}}
\newc{\bR}{{\tilde b_R}}
\newc{\eL}{{\tilde e_L}}
\newc{\eR}{{\tilde e_R}}
\newc{\mhp}{m_{H^\pm}}
\newc{\mhalf}{m_{1/2}}

\newc{\lR}{\tilde{l}_R}
\newc{\lL}{\tilde{l}_L}
\newc{\nL}{\tilde{\nu}_L}
\newc{\na}{\chi^0_1}
\newc{\nb}{\chi^0_2}
\newc{\nc}{\chi^0_3}
\newc{\nd}{\chi^0_4}
\newc{\ca}{\chi^{\pm}_1}
\newc{\cb}{\chi^{\pm}_2}
\newc{\capos}{\chi^{+}_1}
\newc{\caneg}{\chi^{-}_1}

\def\PRD#1#2#3{Phys. Rev. D {\bf #1} (19#2) #3}

\def\beq{\begin{equation}}
\def\eeq{\end{equation}}
\def\bea{\begin{eqnarray}}
\def\eea{\end{eqnarray}}
\def\slashchar#1{\setbox0=\hbox{$#1$}           
   \dimen0=\wd0                                 
   \setbox1=\hbox{/} \dimen1=\wd1               
   \ifdim\dimen0>\dimen1                        
      \rlap{\hbox to \dimen0{\hfil/\hfil}}      
      #1                                        
   \else                                        
      \rlap{\hbox to \dimen1{\hfil$#1$\hfil}}   
      /                                         
   \fi}                                         %
\catcode`@=11
\long\def\@caption#1[#2]#3{\par\addcontentsline{\csname
  ext@#1\endcsname}{#1}{\protect\numberline{\csname
  the#1\endcsname}{\ignorespaces #2}}\begingroup
    \small
    \@parboxrestore
    \@makecaption{\csname fnum@#1\endcsname}{\ignorespaces #3}\par
  \endgroup}
\catcode`@=12

\def\jfig#1#2#3{
 \begin{figure}
 \centering
 \epsfysize=3.25in
 \hspace*{0in}
 \epsffile{#2}
 \caption{#3}
 \label{#1}
 \end{figure}}

\begin{document}

\begin{titlepage}

\begin{flushright}
SLAC-PUB-7148 \\
CERN-TH/96-115 \\
SU-ITP 96-16\\
hep-ph/9604452\\
\end{flushright}


\huge
\begin{center}
 Implications of Low Energy \\Supersymmetry Breaking \\
   at the Tevatron
\end{center}

\large

\vspace{.15in}
\begin{center}

Savas Dimopoulos$^{a,b}$, Scott Thomas$^c$\fnote{\dagger}{Work 
supported by the Department of Energy
under contract DE-AC03-76SF00515.}, James D. Wells$^{c \dagger}$ \\
\vspace{.15in}
\normalsize
$^a${\it Theoretical Physics Division,
CERN\\
CH-1211 Geneva 23, Switzerland}
\\
\vspace{.1in}
$^b${\it Physics Department, 
Stanford University, 
Stanford, CA  94305}\\
\normalsize
\vspace{.1in}
$^c${\it Stanford Linear Accelerator Center,
Stanford, CA 94309\\}

\end{center}
 
 
\vspace{0.15in}

\normalsize

\begin{abstract}



The signatures for low energy supersymmetry breaking
at the Tevatron are investigated.
It is natural that the
lightest standard model superpartner is an electroweak
neutralino, which decays to an essentially massless Goldstino
and photon, possibly within the detector. 
In the simplest models of gauge-mediated supersymmetry
breaking, the production of 
right-handed sleptons, neutralinos, and charginos 
leads to a pair of hard photons accompanied by 
leptons and/or jets with missing transverse energy. 
The relatively hard leptons and softer photons of the 
single $e^+e^- \gamma \gamma + \EmissT$ event observed by CDF
implies this event is best interpreted as arising from left-handed 
slepton pair production. 
In this case the rates for $l^{\pm} \gamma \gamma + \EmissT$ 
and $ \gamma \gamma + \EmissT$
are comparable to that for $l^+l^- \gamma \gamma + \EmissT$.

\end{abstract}

\end{titlepage}

\baselineskip=18pt

\section{Introduction}


If supersymmetry at the electro-weak scale is established,
one of the important questions to be addressed
experimentally is the scale and mechanism of supersymmetry breaking. 
It is often assumed that supersymmetry is broken in a hidden
sector at a very high scale, with the breaking transmitted 
to the visible sector by gravitational
strength interactions. 
It is possible however that supersymmetry is broken at a 
scale not too far above the electro-weak scale, 
with the breaking transmitted by non-gravitational 
interactions \cite{lsgauge,hsgauge}.
In this case the gravitino is naturally the lightest supersymmetric
particle.
The longitudinal component of the gravitino, the Goldstone fermion
of supersymmetry breaking, or Goldstino, $G$, couples to 
ordinary matter through interactions suppressed only by the supersymmetry
breaking scale \cite{fayet}.
This allows the lightest standard model supersymmetric particle
to decay to its partner plus the Goldstino. 
In the simplest models the lightest standard model superpartner is a 
neutralino,
$\chi^0_1$; the dominant decay mode over much of the parameter
space is $\chi^0_1 \to \gamma G$ [3-6].
For a supersymmetry breaking scale below a few thousand TeV this decay
can take place inside the detector. 
Within the context of the 
usual supersymmetric standard model, with 
high scale supersymmetry breaking, radiative decays of neutralinos
are not generic, but
can be achieved by tuning parameters \cite{rad,gordy}.
The presence of two hard photons and missing transverse energy
in the final state is therefore a distinctive and generic
signature for low scale supersymmetry breaking.

At a hadron collider the production rates for supersymmetric states
and subsequent cascade decays  
are determined by both the masses and gauge couplings. 
The form of the superpartner mass spectrum
is determined by the
interactions which transmit supersymmetry breaking to the 
visible sector.
With low scale supersymmetry breaking,
one of the simplest possibilities 
is that these interactions are just the ordinary gauge 
interactions \cite{lsgauge,hsgauge,dn}.
The superpartner masses are then roughly proportional 
to their gauge couplings squared. 
This generally implies that the gluino and squarks are too heavy
to be produced at the Tevatron. 
The largest production rates are 
for sleptons, charginos, and neutralinos.
As discussed below, the relative rates and kinematics 
in the various channels can be sensitive to the 
superpartner mass spectrum, and in turn to details
of the messenger sector in which supersymmetry is broken. 

To illustrate the sensitivity of different channels to the form
of the messenger sector, we consider a number of scenarios
which can arise with gauge mediated supersymmetry breaking,
and identify important generic features of the signatures. 
In the next section the minimal model of gauge
mediation is reviewed. 
In this model, if $\chi_1^0$ is mostly
gaugino, its production is suppressed by the large squark masses. 
Pair production of right-handed sleptons, 
and subsequent cascade decays through
$\na$, leads to the final state $\llgg + \EmissT$ \cite{sbtalk,ddrt}.
In addition, chargino and neutralino pair production 
leads to the final states
$WW \gag + \EmissT$, and  
$W \llgg + \EmissT$ or $WZ \gag + \EmissT$.
In section 3 the minimal model with an approximate $U(1)_R$ symmetry
is considered. 
In this case, the gauginos are lighter than in the minimal model,
leading to relatively larger production rates for 
charginos and neutralinos. 
In section 4 the minimal model is considered in the case in 
which $\na$ is roughly equal mixtures of gaugino and Higgsino.
This gives rise to additional final states including
$ jj \gag + \EmissT$ and 
$\lgg + \EmissT$.
In section 5 models in which left-handed sleptons are
lighter than in the minimal model are considered. 
Pair production of left-handed sleptons gives, in
addition to $\llgg + \EmissT$, the final states
$\lgg + \EmissT$ and $\gag +\EmissT$
at comparable rates. 
For definiteness we assume throughout that the lightest 
standard model superpartner is a neutralino, and that
its decay to a photon plus Goldstino is prompt.
Consequences of relaxing the latter assumption are discussed
in the final section.

A single event of the type $\eegg + \EmissT$ 
has been reported by the CDF collaboration 
\cite{park}. 
Such a signature is consistent with slepton pair production, and 
low scale supersymmetry breaking \cite{ddrt,gordy}.
In section 6 we consider this interpretation of the event
within the context of the models discussed below. 
The kinematics of the event, namely hard leptons and somewhat softer
photons, 
and apparant lack of many other events 
with jets in the final state,
are most easily accommodated with left-handed slepton production. 
In this case, the additional final states mentioned above 
should be seen at comparable rates.

\section{The Minimal Model of Gauge-Mediated Supersymmetry Breaking}

If supersymmetry is broken at a low scale, the ordinary gauge
interactions can act as messengers of supersymmetry breaking. 
The simplest possible messenger sector, 
which preserves the successful prediction
of $\sin^2 \theta_W$ at low energy, are fields which possess the 
quantum numbers of a single $\bf{5} + \bar{\bf{5}}$ of $SU(5)$.
The triplets, $q$ and $\bar{q}$, and doublets $\ell$ and $\bar{\ell}$,
of $\bf{5} + \bar{\bf{5}}$, couple to a single background field, $S$, 
through a superpotential 
$W = S( \lambda_1  q \bar{q} + \lambda_2 \ell \bar{\ell})$.
The field $S$ breaks both $U(1)_R$ and supersymmetry through
its scalar and auxiliary components respectively. 
Integrating out the messenger sector fields gives rise 
radiatively to 
both scalar and gaugino masses. 
The visible sector 
gluino and squarks in this model are heavy enough to be beyond
the reach of the Tevatron. 
The masses of the 
left-handed sleptons, $W$-inos (partners of the $SU(2)_L$ gauge bosons),
right-handed sleptons, and $B$-ino (partner of the $U(1)_Y$ gauge boson),
are in the ratio $2.5~:~2~:~1.1~:~1$.  
We will refer to this model as the Minimal Gauge-Mediated (MGM) model
of supersymmetry breaking. 
The dimensionful terms in the Higgs sector required 
to break the $U(1)_{PQ}$ and $U(1)_{R-PQ}$ symmetries, 
$W=\mu H_1 H_2$ and $V=-m_{12}^2 H_1 H_2 + h.c.$, 
must arise from additional interactions
\cite{dn,mu,dnns}, and may be taken as free
parameters in the minimal model.
Values of $|\mu |$ larger than roughly 150 GeV 
are mildly preferred in order to suppress 
charged Higgs contributions to ${\rm Br}(b \to s \gamma)$ \cite{dnns}.
For the mass ranges considered below, the lightest two neutralinos,
$\na$ and $\nb$,
and lightest chargino $\ca$,
are then mostly gaugino, with small Higgsino mixtures. 
In the $\mu \gg m_{\na}$ 
limit, the spectrum of light states is in the ratios given
above, and the most important parameter which determines the 
phenomenology at the Tevatron is just the overall scale. 

The production rate for the light states depends on both
the masses and charges. 
If the lightest neutralinos are mostly gaugino, 
$\na$ is mostly $B$-ino.
Pair production of $\na \na$ through off-shell $Z^*$ exchange
is then suppressed by a small coupling,
and through $t$- and $u$- channel squark exchange 
by the large squark masses. 
However, 
pair production of $\lR \lR$ through off-shell $\gamma^*$ and $Z^*$,
and subsequent cascade decay 
$\lR \to \l \na$, leads to the final state $\llgg + \EmissT$ 
\cite{sbtalk,ddrt}.
In addition, pair production of charginos
and neutralinos through an off-shell $W^*$ 
(via coupling to the $W$-ino components) leads to 
comparable production rates for $\chi^+_1 \chi^-_1$ and 
$\nb \ca$.
For large $|\mu|$ the neutralino $\nb$ decays predominantly by
$\nb \to \lR l$.
For any reasonable $\mu$ and 
$m_{\ca} > m_{\na} +  m_{W}$, the chargino $\ca$
decays predominantly
through its Higgsino components to the Higgsino components of $\na$
by $\ca \to \na W$. 
On the other hand, for $m_{\ca} < m_{\na} +  m_{W}$, 
$\ca$ decays to three body final
states predominantly through off-shell $W^*$ and $\lR^*$.
The total cross sections which arise at the Tevatron 
in this model with $m_{\chi^0_1} = 100$ GeV and $\mu \gg m_{\chi^0_1}$
are given in Table~1. 
\begin{table}
\centering
\begin{tabular}{lccc}
\hline \hline
             & $\llgg$ & $W\leplep \gag$ & $WW \gag$ \\ \hline
$\lR \lR$ & 6 & - & - \\
$\na \cb$ & - & 11.5 & - \\
$\capos \caneg$ & - & - & 18.8 \\ \hline
Total & 18 & 34.4 & 18.8 \\
\hline \hline
\end{tabular}
\label{tableMGM}
\caption{Production cross sections (fb) for each lepton flavor 
within the MGM for $m_{\chi_1^0} = 100$ GeV, $\mu \gg m_{\chi_1^0}$, 
and $m_{\tilde l_R} =110$ GeV, as discussed in section 2. 
The center of mass energy is 1.8 TeV. Each final 
state has $\not \! \!  E_{T}$. The total cross sections in each channel 
are summed over all lepton flavors.}
\end{table}
In this case $\na$ is pure $B$-ino and $\nb$ and $\ca$ are
pure $W$-ino. 
In the $\sin^2 \theta_W \to 0$ limit 
$\sigma(\nb \ca) = 2 \sigma( \capos \caneg)$.

For finite $\mu$ constructive or destructive interference with 
the Higgsino mixtures in $\nb$ and $\ca$ can significantly
affect the cross sections. 
For example, with $\mu = -250$ GeV, $m_{B} = 100$ GeV, and
$\tan \beta  \equiv \langle H_2 \rangle / \langle H_1 \rangle = 2$,
$\sigma(\nb \ca) \simeq 25.4$ fb and 
$\sigma(\capos \caneg) \simeq 13.7$ fb.
The branching ratios can also be modified for finite $\mu$. 
For the above parameters, $m_{\nb} - m_{\na} > m_Z$ and so
$\nb$ decays predominantly through its Higgsino components to 
the Higgsino components of $\na$ by $\nb \to \na Z$. 
The final states $W \llgg + \EmissT$
are then replaced by $WZ \gag + \EmissT$.

The total rates of course depend on the overall 
scale, but the relative rates in the various channels 
are a slow function of the overall scale. 
The final states $\llgg + \EmissT$, $W \leplep \gag+\EmissT$,
and $WW\gag + \EmissT$, therefore represent an important
test of the MGM in the large $|\mu|$ limit.
The relative rates in the $W \llgg + \EmissT$ and 
$WZ \gag + \EmissT$ are sensitive to the magnitude
of $\mu$, as discussed above. 
In addition, if the usual gauge interactions are the dominant
messengers of supersymmetry breaking, it follows that 
the right-handed sleptons
are essentially degenerate. 
Final states for each lepton flavor should have equal rate.
Because of the relatively large mass of the left-handed sleptons, 
pair production of $\lL \lL$ through off-shell $\gamma^*$ and $Z^*$,
and $\nL \lL$ through off-shell $W^*$,
are suppressed in the MGM. 
For example, with the parameters given in Table~1, 
$\sigma(\nL \lL) / \sigma(\lR \lR) \simeq 0.04$ and
$\sigma(\lL \lL) / \sigma(\lR \lR) \simeq 0.025$.

An important feature of the MGM is the kinematics of the
partons in the final states. 
Since the mass splitting between $\lR$ and the 
$B$-ino is so small, the decay $\lR \to l \na$ results in
fairly soft leptons. 
In contrast, for the decay $\na \to \gamma G$, the photon
receives half the $\na$ mass in the rest frame, resulting
in a larger average photon energy. 
In addition, since $\na$ is generally boosted in the lab frame, the
photon $E_T$ spectrum is much flatter than that of the leptons. 
The $E_T$ and $\EmissT$ for the $\llgg + \EmissT$ final state
with the parameters of Table 1 are shown in Fig. 1 \cite{isajet}.
This illustrates how the kinematics can be used to infer mass
splittings within a decay chain.

\jfig{kineReR}
{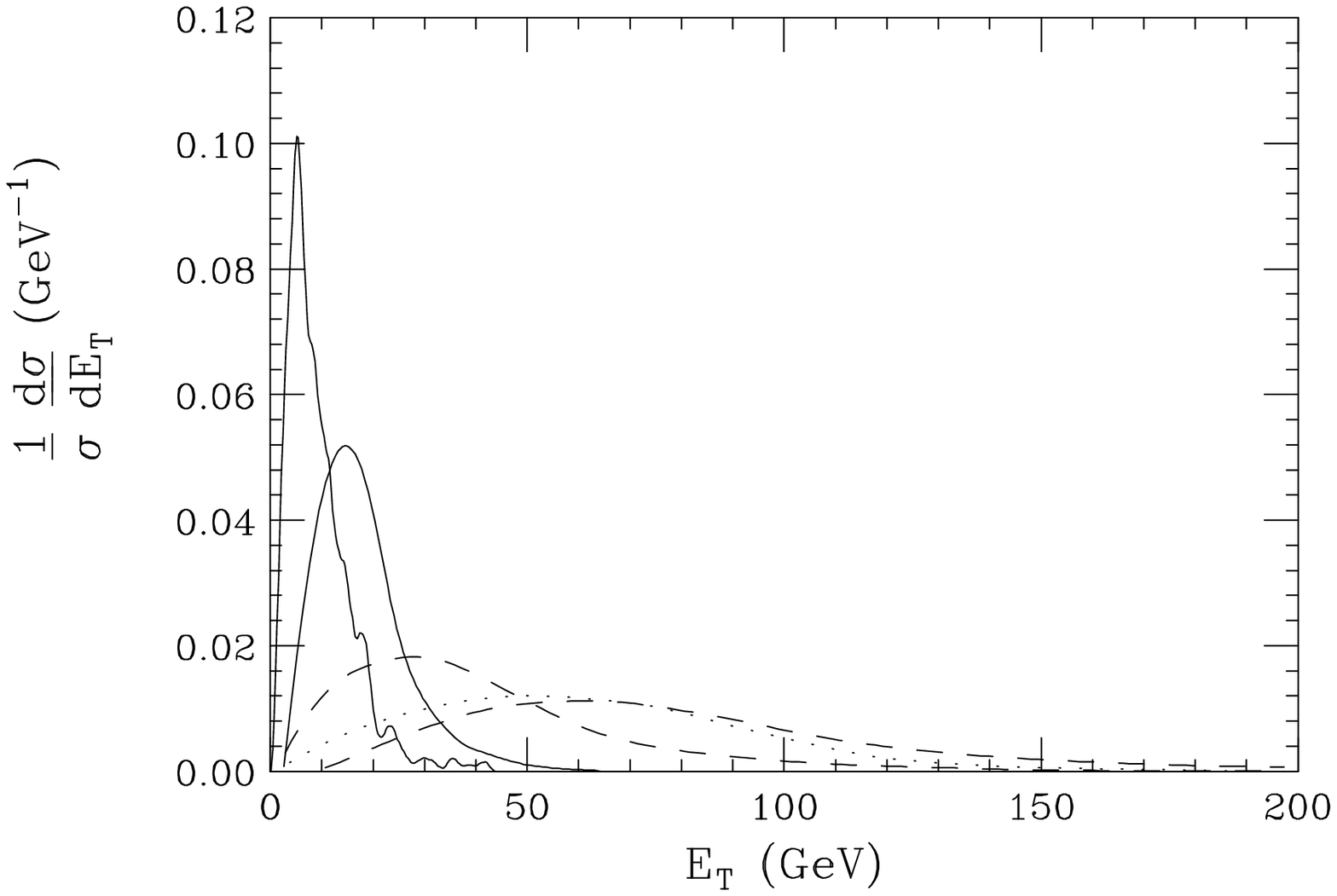}
{The
$E_T$ and $\EmissT$ spectra for the $\llgg + \EmissT$ channel
in the MGM model with the parameters given in Table 1. 
The two solid lines are 
the $E_T$ distributions of the hard and soft electron. 
Similarly, the dashed lines are the $E_T$ distributions of the
hard and soft photon.  The dotted line is the $\EmissT$ distribution.}

\section{Minimal Gauge Mediation with an Approximate $U(1)_R$ Symmetry} 

Scalar masses require supersymmetry breaking, whereas
gaugino masses require the breaking of both supersymmetry and 
$U(1)_R$ symmetry. 
In the MGM a single field, $S$, is assumed to communicate the
breaking of both $U(1)_R$ and supersymmetry to the messenger
sector. 
This is the origin of the relation between the gaugino and scalar 
masses. 
In general, however, these two symmetries can be broken in different
sectors.
As a simple example, consider a messenger sector with fields which 
carry the quantum numbers of two generations of ${\bf 5}+\bar{\bf 5}$,
with superpotential 
$ 
W =  \lambda X ({\bf 5}_1 \bar{\bf 5}_1 + \xi^2)
  +  \lambda^{\prime} S 
         ({\bf 5}_1 \bar{\bf 5}_2 + {\bf 5}_2 
            \bar{\bf 5}_1 )
$.
For $\lambda^{\prime} S > \xi$, ${\bf 5}_i = \bar{ \bf 5}_i =0$, 
and $X$ and $S$ are undetermined at tree level. 
Supersymmetry is broken for $\xi \neq 0$, while a $U(1)_R$ symmetry
is broken for $\lambda X \neq 0$.
For $\lambda X \ll \lambda^{\prime} S$ there is an approximate
$U(1)_R$ symmetry, and the visible sector gauginos can be significantly
lighter than the scalars.

It is possible then that the small mass splitting 
between $\lR$ and the $B$-ino which exists in the MGM is
larger in more general models. 
This has the effect of decreasing the $\lR \lR$ production 
rate relative to the $\capos \caneg$ and $\nb \ca$ rates.
For large enough mass splitting, it is possible that 
$m_{\lR} > m_{\nb} - m_{\na}$. 
The neutralino $\nb$ then decays
predominantly through its Higgsino components to the Higgsino
components of $\na$ by $\nb \to \na Z$. 
The relative rates in the final states 
$\leplep \gag + \EmissT$, $W \leplep \gag + \EmissT$, and $WZ \gag + \EmissT$
are therefore sensitive to the mass splitting between
$\lR$ and the $B$-ino in the the MGM with an approximate $U(1)_R$
symmetry. 
Independent of the $\llgg + \EmissT$ final state, 
if $m_{\lL}, m_{\tilde{\nu}_L} > m_{\nb}$, 
$WW \gag + \EmissT$ and the sum of $W \leplep \gag + \EmissT$ and 
$WZ \gag + \EmissT$ final states represent an important test of whether 
the two lightest neutralinos are mostly gaugino within low scale 
gauge-mediated supersymmetry breaking. 

An important distinction for models with an approximate
$U(1)_R$ symmetry is the kinematics of the partons in 
the $\llgg + \EmissT$ final states. 
Since the mass splitting between $\lR$ and the $B$-ino
can be larger than in the MGM, the decay 
$\lR \to l \na$ gives rise to harder leptons. 
Even with a relatively small number of events it should be
possible to distinguish between models with the MGM mass relations, 
and more general models with larger mass splittings. 


\section{Minimal Gauge Mediation with Higgsino Production}

The four neutralinos of the minimal supersymmetric standard model
are in general a mixture of gauginos and Higgsinos. 
For $\mu$ comparable to the gaugino masses, 
pair production of neutralino and charginos through the Higgsino
components can give rise to additional important channels. 
An example which illustrates the case in which the lightest
neutralinos are roughly equal mixtures of gaugino and Higgsino
are for the parameters $\mu=-160$ GeV, $m_{B} = 150$ GeV,
and $\tan \beta =2$.
This choice of $\mu$ represents a low value which is still 
marginally consistent with ${\rm Br}(b \to s \gamma)$ \cite{dnns}.
The neutralino mass eigenvalues are then 
144, 169, 177, and 322 GeV.
The two lightest neutralinos, $\na$ and $\nb$, are roughly equal mixtures of 
$B$-ino and the symmetric combination of Higgsinos.
The neutralino $\nc$ is 
mostly the anti-symmetric combination of Higgsinos, while $\nd$ is 
mostly $W$-ino.
The coupling of an off-shell $Z^*$ to pairs of 
nearly symmetric or anti-symmetric 
Higgsinos is suppressed, but the coupling of a $Z^*$ to a symmetric
Higgsino and anti-symmetric Higgsino is unsuppressed.
The dominant neutralino pair production with the above parameters
is therefore for $\na \nc$.
Since $m_{\nb} > m_{\lR}$,
the neutralinos $\nb$ and $\nc$ decay predominantly by $\chi_i^0 \to \lR l$.
Production of $\na \nc$ therefore gives the final state
$\llgg + \EmissT$.
In more general models with $m_{\lR} > m_{\nb}$ the neutralinos
decay predominantly to three body final states through off-shell
$\lR^*$ and $Z^*$.

The chargino $\ca$ is mostly Higgsino,
while $\cb$ is mostly $W$-ino, with masses 168 and 322 GeV 
respectively. 
The chargino $\ca$ decays predominantly to three body final 
states through an off-shell $W^*$. 
Pair production of $\capos \caneg$ through off-shell $\gamma^*$ and 
$Z^*$ then leads to $X \gag + \EmissT$ final states, where 
$X = l^+l^-, l^{\pm}jj$, and $4j$.
In addition, production of $\ca \chi_i^0$, $i=1,2,3$, through an off-shell 
$W^*$, leads to the final states $X \gag + \EmissT$, where
$X = l^{\pm}, l^+l^-l^{\prime \pm}, l^+l^-jj$, and $jj$. 
The production cross sections for this set of parameters are
summarized in Table~2.
\begin{table}
\centering
\begin{tabular}{lcccccccc}
\hline \hline
 & $l^{\pm}\gag$ & $l^+l^-\gag$ & $l^+l^-l^{\pm} \gag$ & $\lllgg$ & 
   $l^{\pm}jj\gag$ & $l^+l^-jj\gag$ & $jj\gag$ & $4j\gag$ \\
\hline
$\lR \lR$ & - & 1.0 & - &  - & - & - & - & - \\
$\na \nc$ & - & 4.3 & - & - & - & - & - & - \\
$\capos \caneg$ & - & - & - &  - & 1.9 & - & - & 5.7 \\
$\ca \na$ & 2.3 & - & - & - & - & -  & 14.0 & - \\
$\ca \nb$ & - & - & 0.5 & 0.5 &  - & 3.0 & - & - \\
$\ca \nc$ & - & - & 0.9 & 0.9 &- & 5.1 & - & - \\
\hline
 Total & 6.9 & 15.9 & 4.2 & 8.4 & 5.7 & 24.3 & 14.0 & 5.7 \\
\hline \hline
\end{tabular}
\label{tablemu}
\caption{Production cross sections (fb) 
for each lepton flavor 
within the MGM 
for $\mu = -160$ GeV, 
$m_{\tilde B} = 150$ GeV, $m_{\tilde e_R} = 165$ GeV, 
as discussed in section 4. 
The center of mass energy is 1.8 TeV.
Each final state has $\not \! \!  E_{T}$.
The total cross sections in each channel
are summed over all lepton flavors.}
\end{table}
The mostly $W$-ino states $\nd$ and $\cb$ are too heavy to have
appreciable rates with these parameters.

One feature from Table~2 which generically
distinguishes production through Higgsino components 
is the relatively large rate for $\ca \na$.
This leads to the final states $\jjgg + \EmissT$ and $\lgg + \EmissT$,
which do not occur in the mostly gaugino scenario. 
As another example, for the parameters  given in Table 1
with $\mu = 250$ GeV, the Higgsino components of $\na$ and $\ca$
give 
$\sigma(\na \ca) \simeq 40$ fb. 
In addition, the other cross sections are increased to 
$\sigma( \nb \ca) \simeq 105$ fb, and 
$\sigma(\capos \caneg) \simeq 54$ fb.

An additional feature for production of neutralinos and 
charginos with large Higgsino fractions are the kinematics of the final 
states. 
Since Higgsino masses are determined by one mass parameter ($\mu$)
they tend to be fairly degenerate,
whereas the gauginos are more split
($M_2 \simeq  2 M_1$).  
For $\mu \sim M_1$ this leads to mass splittings among the 
light neutralinos which are much smaller than the overall scale.  
Thus,
the leptonic and/or hadronic activity coming from cascade decays
down to $\chi^0_1$ tends to be much softer than the photons arising
from $\chi^0_1\to \gamma G$.

\section{Non-minimal Models with Left Handed Slepton Production}

The ratio $m_{\lL} / m_{\lR} \simeq 2.3$ is fixed in the MGM by the 
form of the messenger sector and the relative magnitude of the 
$SU(2)_L$ and $U(1)_Y$ gauge couplings. 
In more general models this ratio can be modified. 
For example, additional gauge interactions under which
$\lR$ is charged, or direct Yukawa couplings of the messengers with the
Higgs or matter multiplets, can in general reduce this ratio. 
Pair production of $\tilde l_L\tilde l_L$, $\tilde l_L\tilde \nu_L$
and $\tilde \nu_L\tilde \nu_L$ can dominate $\tilde l_R\tilde l_R$ even
for $m_{\tilde l_L},m_{\tilde \nu_L}>m_{\tilde l_R}$ because of the
larger $SU(2)_L$ gauge coupling for left-handed states.
For $m_{\lL}, m_{\nL} < m_{\nb}$
the left handed sleptons and sneutrinos decay predominantly 
by $\lL \to \na l$ and $\nL \to \na \nu$,
leading to the final states $\llgg + \EmissT$, 
$\lgg + \EmissT$, and $\gag + \EmissT$
at comparable rates. 
In addition, for $m_{\lL}, m_{\nL} < m_{\nb}$,
$\nb$ and $\ca$ decay predominantly by 
$\nb \to \lL l, \nL \nu$ and 
$\ca \to \tilde{l}_L^{\pm} \nu, \nL l^{\pm}$.
Pair production of $\chi^+ \chi^-$ then leads to the final states
$l^+ l^{\prime -} \gag + \EmissT$, and 
$\nb \ca$ leads to 
$l^{\pm} \gag + \EmissT$ and $l^+ l^- l^{\prime \pm} \gag + \EmissT$
final states.
To illustrate the features of such a spectrum,
the production cross sections for 
$m_{\lL} = 135$ GeV, $m_{\nL}=120$ GeV, and $m_{\lR} = 110$ GeV
are presented in Table~3.
%
\begin{table}
\centering
\begin{tabular}{lcccccc}
\hline \hline
             & $\gag$ & $\lgg$ & $\llgg$ & 
               $l^+l^{\prime -}\gag$ & $l^+l^-l^{\pm} \gag$ & $\lllgg$ 
     \\ \hline
$\lL \lL$ & - &  - & 5.6 & - & - & -\\ 
$\lL \nL$ & - & 17.0 & - & - & - & - \\
$\nL \nL$ & 6.7 & - & - & - & -  & -\\
$\lR \lR$ & - & - & 6.0 & - & - & -\\
$\nb \ca$ & - & 6.5 & - & - & 1.65 & 1.65 \\
$\capos \caneg$ &-  &- & 2.1 & 2.1  & - & -\\ \hline
Total & 20.0 & 70.5 & 41.1  & 12.6 & 5.0 & 9.9 \\
\hline \hline
\end{tabular}
\label{tablelL}
\caption{Production cross sections (fb) 
for each lepton flavor 
for $m_{\chi_1^0} = 100$ GeV, $\mu \gg m_{\chi_1^0}$, 
$m_{\lR} = 110$ GeV, $m_{\lL} = 135$ GeV, $m_{\nL}=120\gev$,
as discussed in section 5.
The center of mass energy is 1.8 TeV. 
Each final state has $\EmissT$.
The total cross sections in each channel are summed over all 
lepton flavors.}
\end{table}
The $\lL - \nL$ mass splitting is that which arises from 
the $SU(2)_L \times U(1)_Y$ $D$-terms,
$m_{\lL}^2 - m_{\nL}^2 = -m_W^2 \cos 2 \beta$,
for $\tan \beta =2$. 
The lightest neutralinos are taken to be mostly gaugino, 
and the mass ratios of the right-handed sleptons
to gauginos are taken to be those of the MGM. 
It is interesting to note that with this spectrum, no jets
result from the cascade decays.

Left-handed slepton pair production gives rise to very distinctive
final states. 
In the $\sin^2 \theta_W \to 0$ and $\tan \beta \to 1$ limit,
$\sigma(\lL \nL) = 2 \sigma(\lL \lL) = 2 \sigma(\nL \nL)$.
Final states $l^{\pm} \gag +\EmissT$ and $\gag +\EmissT$ in roughly
this ratio represent an important test for $\lL \lL$ production. 
The relative rate
of $l^+l^-\gag +\EmissT$ 
events depends on 
the mass of the 
right-handed slepton with respect to the left-handed slepton.
A rate for $\lL \lL$ comparable to or larger than that for 
$\lR \lR$ would imply a mass spectrum which is not consistent 
with the MGM mass relations.

\section{Implications of Current Data} 

It is by now well-known that 
a single event of the type $\eegg + \EmissT$ 
has reportedly been observed by the CDF collaboration 
\cite{park}. 
The single event is consistent with $\tilde{e}^+ \tilde{e}^-$ pair
production, and subsequent decay 
$\tilde{e} \to \na e$ and $\na \to \gamma G$,
within low scale supersymmetry breaking \cite{ddrt,gordy}.
For a single event in 
$\sim 100$ pb$^{-1}$ of integrated luminosity, the 90\% CL 
range for the cross section is roughly  
5 $-$ 40 fb. 
The kinematics of the event requires
$m_{\tilde{l}} \gsim 60$ GeV. 
If $\na$ is mostly gaugino, the non-observation of 
an excess in $e^+e^- \to \gag + \EmissT$ at LEP135 \cite{LEPgg}
gives a bound on the $\na$ mass almost to kinematic
threshold, $m_{\na} > 65$ GeV. 
Using this, the kinematics of the $\eegg + \EmissT$ 
event require $m_{\tilde{l}} \gsim 90$ GeV.
If $e^+e^- \to \gag + \EmissT$ were not observed
at LEP190, this would increase to
$m_{\tilde{l}} \gsim 110$ GeV. 
Given the analysis of the previous sections, it is interesting
to investigate what consequences the Goldstino
interpretation of this event has for the messenger sector, 
and for other channels which could be observed at the Tevatron
and LEPII.

Within the MGM model, the most natural interpretation 
would be $\lR \lR$ pair production.
Based on the $\lR \lR$ cross section, this would
imply a 90\% CL range for $m_{\lR}$ of  
$70 \lsim m_{\lR} \lsim 115$ GeV,
consistent with the kinematic bounds given above.
This interpretation is somewhat problematic for a number of reasons.
All the partons in the event are fairly hard, 
$E_{T,{e_1}} \simeq 64$ GeV,
$E_{T,{e_2}} \simeq 34$ GeV,
$E_{T,{\gamma_1}} \simeq 32$ GeV, and 
$E_{T,{\gamma_2}} \simeq 40$ GeV.
However, as discussed in section 2, within the MGM 
the leptons in such events should be much softer on average
than the photons. 
For the parameters given in Table~1 the probability 
that both leptons have $E_T > 30$ GeV is $\lsim 2 \%$. 
It is possible for the $m_{\lR} - m_{\na}$ splitting to be larger
than the MGM relation, as discussed in section 3, thereby 
increasing the average lepton $E_T$. 
Right-handed sleptons much heavier than 115 GeV are 
however disfavored by
the implied rate. 
Values of $m_{\na}$ much smaller than 100 GeV are disfavored if
$\na$ is mostly gaugino since $\nb \ca$ and $\capos \caneg$ pair
production would lead to an excessive rate for the final states
$WW \gag + \EmissT$ and $W \leplep + \EmissT$ or 
$WZ \gag + \EmissT$.
We therefore conclude that the kinematics of the event are not
easily accomodated by $\lR \lR$ production if $\na$ is an 
electroweak neutralino.

It is worth noting in passing that it is possible 
for neutralino pair production to result in 
$\llgg + \EmissT$ over some range of parameters,
as discussed in section 4. 
However, the kinematics, and many other concomitant final states 
also disfavor this interpretation.

The problem of obtaining hard leptons is largely
ameliorated if the event is interpreted as arising
from $\lL \lL$ pair production. 
The larger gauge coupling of the left-handed sleptons relative
to the right-handed ones results in a larger intrinsic cross section. 
Based on the $\lL \lL$ cross section, the $90\%$ CL range 
for $m_{\lL}$ is $85 \lsim m_{\lL} \lsim 135$ GeV.
This allows for a larger $m_{\lL} - m_{\na}$ splitting, resulting
in harder leptons. 
As an example, the $E_T$ and $\EmissT$ spectra for 
$m_{\na}=100$ GeV and $m_{\lL} = 135$ GeV are shown 
in Fig. \ref{kineLeL}.
For these parameters the photon and lepton average $E_T$ are
of the same order. 

This interpretation has a number of interesting consequences. 
First, as discussed in section 5, left-handed slepton and sneutrino 
production gives
rise to the additional final states $\lgg + \EmissT$ and 
$\gag + \EmissT$.  
The rate for these final states should be roughly in the 
ratio 2:1 (depending precisely on the value of $\tan \beta$)
and comparable to the $\llgg + \EmissT$ rate. 
In addition, if $\na$ is mostly gaugino
and $m_{\lL}, m_{\nL} < m_{\nb}$,
$\nb \ca$ and $\capos \caneg$ production should give rise to 
final states $X \gag + \EmissT$, where 
$X = l^{\pm}, l^+ l^{\prime -}, l^+ l^- l^{\prime \pm}$ 
at slightly reduced rates. 
However, if $\na$ were mostly electroweak singlet, the rate
for these final states would be suppressed. 

\jfig{kineLeL}
{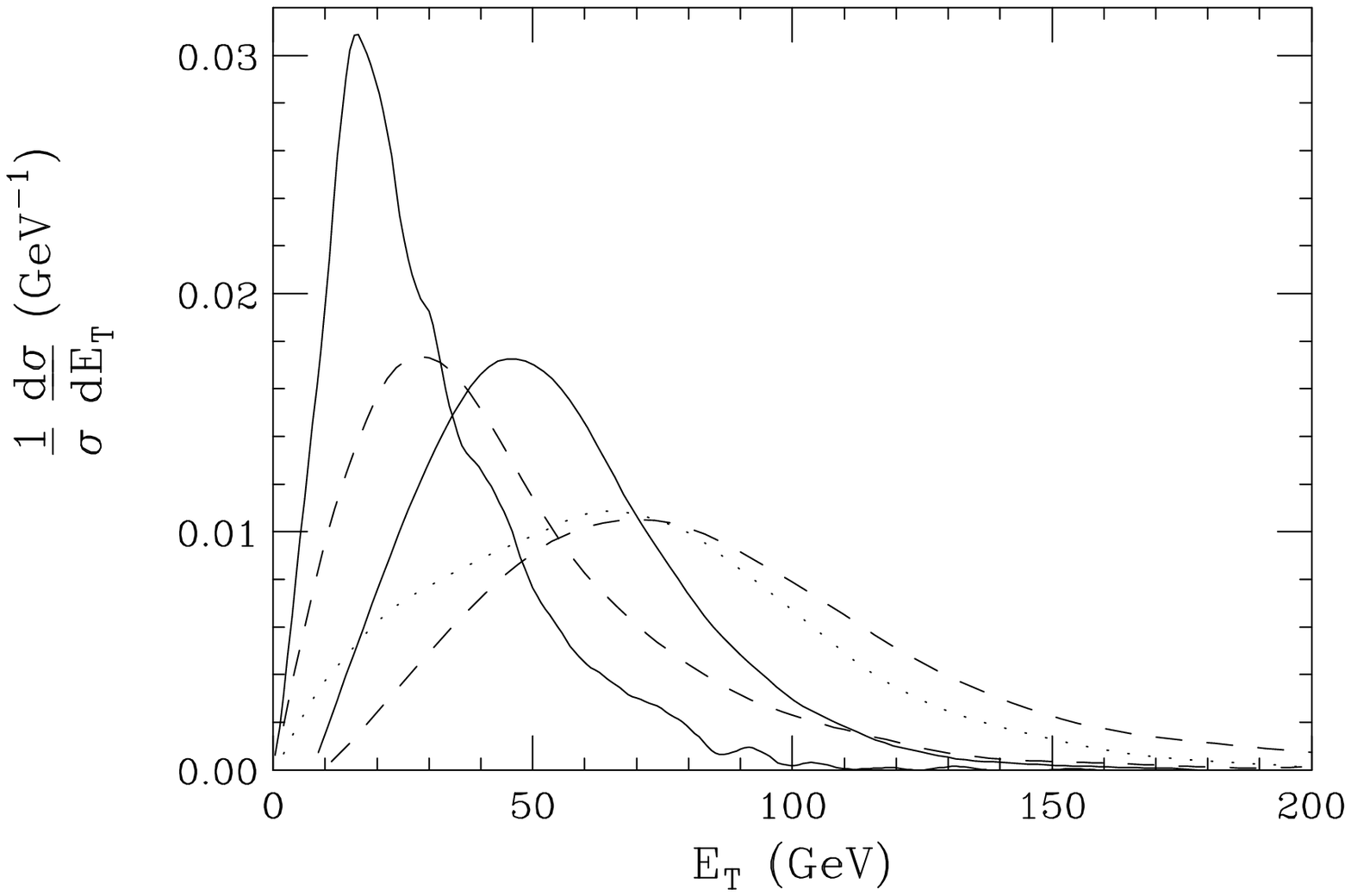}
{The
$E_T$ and $\EmissT$ spectra for the $\llgg + \EmissT$ final state
with $m_{\tilde{l}} = 135$ GeV, $m_{\na} = 100$ GeV,
as discussed in section 6.
The center of mass energy is 1.8 TeV.
The two solid lines are 
the $E_T$ distributions of the hard and soft electron. 
Similarly, the dashed lines are the the $E_T$ distributions of the
hard and soft photon. 
The dotted line is the $\EmissT$ distribution.}

\section{Conclusions}

If supersymmetry is broken at a low scale, the lightest 
standard model supersymmetric particle can decay to its partner
plus the Goldstino. 
This leads to the possibility of signatures which are quite distinct
from those of high scale supersymmetry breaking \cite{ddrt}. 
Here we investigated the phenomenology at the Tevatron for the case 
in which the lightest standard model superpartner is a neutralino. 
The generic feature for this case is a final state with 
two hard photons and missing transverse energy. 
The additional partons in the final state are sensitive 
to the superpartner mass spectrum, and can depend indirectly on details
of the messenger sector.
An observation of $\llgg + \EmissT$ alone 
with soft leptons and hard photons
would be a good
indication of right-handed slepton pair production.
The final states $WW \gag + \EmissT$ and the sum of 
$W \llgg + \EmissT$ and $WZ \gag + \EmissT$ 
in approximately a 2:1 ratio result if the 
lightest two neutralinos and lightest chargino are mostly
gaugino
and $m_{\lL}, m_{\nL} > m_{\nb}$.
The combination of the additional final states $jj \gag + \EmissT$ and 
$\lgg + \EmissT$ can arise if the lightest neutralino has
a significant Higgsino component. 
Finally, the final states $\lgg + \EmissT$, 
$\llgg + \EmissT$, and $\gag + \EmissT$ 
arise in approximately a 2:1:1 ratio from left handed slepton production. 
If $m_{\lL}, m_{\nL} < m_{\nb}$ the additional final
states 
$l^+ l^{\prime -} \gag + \EmissT$ and 
$l^+ l^- l^{\prime \pm} \gag + \EmissT$ are also significant.
If the ordinary gauge interactions are the messengers for supersymmetry
breaking, all final states discussed in this paper will
occur with equal rates for each generation. 
Violations of lepton universality in two photon events
would likely indicate a much richer family-dependent 
messenger sector.  

The kinematics of the above final states are sensitive to the 
mass spectrum. 
Since the photons arise at the end of the decay chain, their
$E_T$ spectrum is generally flatter than for the other partons. 
For very massive superpartners the splitting between states
is typically smaller than the overall scale, giving rise 
to an average photon $E_T$ much larger than for the other partons,
which originate further up the decay chain. 
For superpartners which could be observed with 
the current integrated luminosity at the Tevatron, this is 
however not necessarily the case (cf. Fig. 2). 

With current or 
anticipated integrated luminosities, the 
signatures \hbox{$X \gag + \EmissT$} discussed in this paper have
very small standard model backgrounds. 
The largest potential backgrounds are probably mis-identifications. 
The most problematic of these could be hard $\pi^0 \pi^0$ pairs
which are interpreted as $\gag$. 
However, in the final states with leptonic activity only, this requires
that two hard jets each fluctuate to a single $\pi^0$ plus 
hadronic activity below the pedastool. 
This doubly rare fluctuation could be estimated if 
the single rare fluctuation rate can be characterized experimentally. 

Given these very distinctive final states with negligible standard 
model backgrounds, it is 
interesting to ask what the discovery reach will be at future
hadron colliders. 
For a 5 event signal at the Tevatron with $\sqrt{s}=2$ TeV,
pair production of 
$\lR \lR$ and $\nb \ca$ give mass reaches of 
$m_{\lR} \lsim 145~(220)$ GeV and $m_{\ca} \lsim 300~(380)$ GeV for
an integrated luminosity of 2 (20) fb$^{-1}$.
Notice that these are well beyond the reach of LEPII.
Since the right-handed sleptons are generally lighter
than the chargino, 
both processes probe
roughly the same overall scale. 
The analogous processes at the LHC with $\sqrt{s}=14$ TeV
give mass reaches 
$m_{\lR} \lsim 540$ GeV and $m_{\ca} \lsim 1200$ GeV for
an integrated luminosity of 30 fb$^{-1}$.
In contrast, the reach for the 
often discussed case of high scale supersymmetry 
breaking is significantly lower because of standard model 
backgrounds \cite{trilepton,SUSYhadron}.

Throughout, we have assumed that $\na$ decays promptly 
by $\na \to \gamma G$.
However, for a supersymmetry breaking scale of a few thousand TeV
this decay length can be on the same scale as the detector 
dimensions \cite{ddrt}.
With some fraction of the decays taking place outside the
detector, some events could appear with a single photon, 
or without photons. 
In addition, if $\na$ has a non-negligible Higgsino component,
the decay 
$\na \to h^0 G$ can arise, where $h^0$ is the lightest
Higgs boson.
Some fraction of the events would then have one or both
photons replaced by $bb$ jets reconstructing the Higgs mass. 
This would represent a very interesting, and relatively clean, 
source for Higgs bosons.

The single $\eegg + \EmissT$ event observed at the Tevatron
by the CDF collaboration \cite{park} 
is most naturally
interpreted as low scale supersymmetry breaking, with the 
missing energy carried by Goldstinos. 
The relatively hard leptons and softer photons,
and lack of many other events in other channels, 
suggests $\lL \lL$ pair production as the origin
of this event. 
In this promising scenario $\lL \nL$ and $\nL \nL$ pair production 
gives rise to $\lgg +\EmissT$ and \hbox{$\gag + \EmissT$} final states
at comparable rates. 
If these events are not seen after a complete analysis of the 
current CDF and
D0 data, this interpretation would be somewhat 
problematic.  
In addition, 
for $m_{\lL}, m_{\nL} > m_{\nb}$, 
the final states $WW \gag + \EmissT$ 
and either $WZ \gag + \EmissT$ or $W \llgg + \EmissT$ 
can test the gaugino fraction of $\na$. 
Alternately, if $m_{\lL}, m_{\nL} < m_{\nb}$ the final states
$l^+ l^{\prime -} \gag + \EmissT$ and 
$l^+ l^- l^{\prime \pm} \gag + \EmissT$ can test the
gaugino fraction of $\na$.
If $\na$ is mostly gaugino such signatures are likely to be seen 
in the current data. 
In contrast, if $\na$ is mostly singlet, the rate for 
these final states would be reduced. 

Finally, it is worth commenting on the implications of this
interpretation of the CDF event for LEPII.
Our analysis indicates that slepton pair production 
is likely to be out of reach at LEPII.
However, neutralino pair production is not necessarily out
of reach. 
Its signature would be spectacular $\gag + \EmissT$ events with
acoplanar photons.

{\it Acknowledgements.} We would like to thank M. Dine and H. Haber for 
useful discussions.

\end{document}